\documentstyle[12pt]{article}
\input epsf.sty
\newcommand{\be}{\begin{equation}}
\newcommand{\ee}{\end{equation}}
\newcommand{\bea}{\begin{eqnarray}}
\newcommand{\eea}{\end{eqnarray}}
\newcommand{\al}{\alpha}

\newcommand{\Dl}{\Delta}
\newcommand{\eps}{\epsilon}
\newcommand{\zt}{\zeta}

\newcommand{\kp}{\kappa}
\newcommand{\lm}{\lambda}
\newcommand{\Lm}{\Lambda}

\newcommand{\nn}{\nonumber}

\begin{document}
\title{Dilaton Brane Cosmology with Second Order
String Corrections and the Cosmological Constant}
\author{Apostolos Kuiroukidis\thanks{E-mail:
kouirouki@astro.auth.gr}\\
Department of Physics, \\
\small Section of Astrophysics, Astronomy and Mechanics, \\
\small Aristotle University of Thessaloniki, \\
\small 541 24 Thessaloniki, GREECE\\
}

\maketitle
\begin{abstract}
We consider, in five dimensions, the effective
action from heterotic string which includes quantum gravity corrections
up to $(\al ^{'})^{2}$. The expansion, in the string frame, is in terms
of $|\al ^{'}R|$, where $R$ is the scalar curvature and uses the third
order Euler density, next to the Gauss-Bonnet term. For a positive tension
brane and infinite extra dimension, the logarithmic class of solutions is
less dependent from fine-tuning than in previous formulations. More
importantly, the model suggests that in the full non-perturbative
formulation, the string scale can be much lower
than the effective Planck mass, without the string coupling to be vanishingly
small. Also, a less severe fine-tuning of the brane tension is needed.\\

\end{abstract}

\newpage

\section{Introduction}\

Recent developments in string theory suggest that matter and
gauge interactions may be confined on a brane, embedded in a
higher dimensional space (bulk), while gravity can propagate
into the bulk (for reviews see \cite{epap}, \cite{maar1}).
Within this context several toy models
have been constructed to address such issues as the hierarchy
and cosmological constant problems \cite{aluk}.
In particular, the large hierarchy between the Standard Model and
Planck scales could be explained for an observer on a negative
tension flat brane, if the extra dimension was taken to be compact
\cite{rand1}. The possibility of a large non-compact dimension
was realized in \cite{nark}, while it was shown in \cite{rand2}
that warping of five-dimensional space could lead to localization
of gravity on the brane, even though the size of the extra dimension
was of infinite proper length.

In \cite{arka}, \cite{kach} a simple, interesting alternative model
has been considered, where a bulk scalar field $\phi $ is coupled to
the brane tension $T_{br}$. This is the all-loop contribution to the
vacuum energy density of the brane, from the Standard Model fields.
For the 4D cosmological constant problem considered there, solutions
of the field equations were found, which localize gravity, but possess
naked singularities at finite proper distance. This proper distance
is given by $y_{c}=1/\kp _{(5)}^{2}T_{br}$ where the five and
four-dimensional Planck scales $k_{(5)}^{2}=M_{(5)}^{-3}$,
$\kp _{(4)}^{2}=M_{(4)}^{-2}$ are related by
\be
T_{br}=\frac{\kp _{(4)}^{2}}{\kp _{(5)}^{4}}=
\frac{M_{(5)}^{6}}{M_{(4)}^{2}}.
\ee
Then if we momentarily identify the 4D cosmological term with the
brane tension $\lm =T_{br}\sim (1TeV)^{4}=(10^{3}GeV)^{4}\sim 10^{-64}
M_{(4)}^{4}$ we obtain
\be
M_{(5)}\simeq 10^{8}GeV,\; \; y_{c}\simeq 1mm,
\ee
which is acceptable by present day experiments.

It was soon realized that the bulk action should also contain the
{\it classical} Gauss-Bonnet (GB) term
\be
{\cal L}_{GB}=R^{2}-4R_{ab}R^{ab}+R_{abcd}R^{abcd},
\ee
which is the
leading quantum gravity correction, and the only to provide second
order field equations. Some of the early works on the GB gravity
include \cite{egb}, while the corresponding brane cosmology has
been studied, among others, in \cite{rgb}. The corresponding
generalized junction conditions appeared in \cite{jun}.

The GB term is the first in an infinite series of Euler densities
that appear in the generalized Lovelock Lagrangian \cite{lov}.
The next order term is the third order, given by
\bea
{\cal L}_{(3)}&=&2R^{abcd}R_{cdef}R^{ef}\; _{ab}
+8R^{ab}\; _{ce}R^{cd}\; _{bf}R^{ef}\; _{ad}
+24R^{abcd}R_{cdbe}R^{e}\; _{a}+\nn \\
&+&3RR^{abcd}R_{cdab}
+24R^{acbd}R_{ba}R_{dc}+16R^{ab}R_{bc}R^{c}\; _{a}
-12RR^{ab}R_{ba}+R^{3}.
\eea
It is the only combination of three contractions of the Riemann tensor,
that is linear in the second derivatives of the metric tensor
\cite{fmull} and linearization about flat Minkowski space introduces
no propagator corrections (i.e., no ghosts) \cite{zw}.
Discussion of cosmological problems with cubic and/or quartic
Lagrangians, in the framework of superstring theories, appeared in
\cite{fmcos}. Early work on global symmetries of higher order gravity
theories appeared in \cite{meiss} while
solutions without bulk cosmological constant appeared in \cite{meiss1}.

The effective gravitational equations on the brane have been studied in
\cite{shir1}, while the inclusion of the induced gravity model has
been studied, among others, in \cite{maed1}. The effective equations
on the brane, in the presence of the GB contribution in the bulk,
were studied in \cite{maed2}, whereas astrophysical implications
of higher order gravity models on branes have been analyzed in
\cite{lik}. The cosmological perturbations in brane models have been
examined in \cite{maar}, while the short scale modifications of Newton's
law have been dealed with in \cite{kofin}.

In \cite{ftun} it was shown that in general self-tuning is generic
in theories with at most two branes, or a single brane with orbifold
boundary conditions. Also it was shown that localization of gravity
(namely finiteness of the effective Planck mass) and infinite extra
dimension is only consistent with fine-tuning of the parameters of
the theory. A no-go theorem appeared in \cite{dasg}. The problem
of fine-tuning (i.e., the adjustment of Lagrangian parameters
as for example the relation of bulk and brane cosmological constants
in RS models) was shown not to be present in \cite{jak1}. Solutions
and stability of them which are potentially self-tuning (i.e., need for
adjustment of integration constants rather than Lagrangian parameters,
as for example the values of the potential $V(\phi _{0})$ and its derivative
$V^{'}(\phi _{0})$ on the brane) where shown to exist.
The issue of self-tuning was studied, for example, in \cite{self}.

In this paper we consider the tree-level effective action from
heterotic string theory, in the string frame (Eq. (5) below).
This has been derived in [28], [31], [32] for a generic spacetime dimension
$D$ and it preserves the general symmetries of the underlying theory
such as general covariance. In the critical dimension $D=10$ the bulk
cosmological constant vanishes. This action has been used by a
number of authors ([29], [30], [33], [34]) also for $D=5$ to study braneworld
models. This is in contrast to the usual models, where in the "Einstein frame"
the higher order corrections (such as the Gauss-Bonnet term) enter without the
exponential dilaton dependence. This form besides the fact that it
stems from string theory, offers a new way to examine "self-tuning" mechanisms
([27], [29]). So we assume that the higher order
$(\al ^{'})$-corrections enter in this action in powers of $|\al ^{'}R|$,
where $R$ is the scalar curvature [29]. This is reasonable because, in the
absence (for simplicity) of gauge field contributions the only length scale of the
theory is $\al ^{'}=l_{s}^{2}$. Exploring this possibility we show that
\begin{itemize}
\item It is free from naked singularities, the curvature scalar being regular
everywhere. Also the effective Planck mass calculated is finite, while the extra
dimension can be infinite.
\item The solutions do not have a smooth limit as the higher order terms are set
to zero, so these are genuine ${\cal O}(\al ^{'2})-$classes of solutions, and more
importantly,
\item The string scale is much lower than the effective Planck mass, the string
coupling assumes higher values, as higher order terms are included, and this
is true for a continuous range of parameters, namely the need for fine-tuning is
much less severe.
\end{itemize}
Our line of reasoning and conventions follow mainly those of
\cite{binet}. All calculations are lengthy and have been performed
with great care.
\section{The Action }\
The action in the string frame, from the string tree level
effective action, must have the same dilaton dependence on the
Einstein-Hilbert, Gauss-Bonnet and third order terms with
respect to the curvature. This is also true due to
dimensional arguments because the string expansion parameter
$\al ^{'}=l_{s}^{2}=M_{s}^{-2}$ is the only length scale of the
theory and quantum gravity corrections appear in this action as
powers of $|\al ^{'}R|$ \cite{binet}.
So, in order to explore the results from this point of view
and ignoring any gauge field contributions for simplicity,
we must have (\cite{mets}, \cite{gross}, \cite{char1})
\bea
S_{bulk}^{(string)}=\frac{M_{s}^{D-2}}{2}\int d^{D}x
\sqrt{-\bar{g}}e^{-2\Phi }
\left\{\frac{}{}\bar{R}+4(\nabla \Phi )^{2}\right.
+\lm _{0}(\al ^{'})[\bar{{\cal L}}_{GB}+...]+\nn \\
+\lm _{1}(\al ^{'})^{2}[\bar{{\cal L}}_{(3)}+...]
\left.-\frac{2(D-10)}{3\al ^{'}}+{\cal O}(\al ^{'3})\right\}.
\eea
Here $\lm _{0}=1/4,1/8,0$ for the bosonic, heterotic and
type II theories \cite{frad}, \cite{mets}. The constant $\lm _{1}$
would in principle stem from string theory and the full contribution
of the third order term is with a value of the order of unity.\\
One must apply a conformal transformation
$\bar{g}_{\mu \nu }=e^{\zt \Phi }g_{\mu \nu }$,
to bring the action in the Einstein frame,
where $\zt =4/(D-2)$ and the metric $g_{\mu \nu }$ is
\be
ds^{2}=e^{2A(y)}\eta _{\mu \nu }dx^{\mu }dx^{\nu }+dy^{2},
\ee
which assures four-dimensional Poincare invariance.\\
For the metric of Eq. (6) we have
$R=-8A^{''}-20(A^{'})^{2}$ for the scalar curvature,\\
${\cal L}_{GB}=24(A^{'})^{2}[4A^{''}+5(A^{'})^{2}]$ for
the Gauss-Bonnet term and\\
${\cal L}_{(3)}=-24(A^{'})^{4}[4A^{''}+15(A^{'})^{2}]$
for the third order term. The corresponding quantities in the
string (barred) frame are given by
\bea
\bar{R}=e^{-\zt \Phi }[-8A^{''}-4\zt \Phi ^{''}
-16\zt A^{'}\Phi ^{'}-3\zt ^{2}(\Phi ^{'})^{2}-20(A^{'})^{2}]
\nn \\
\bar{{\cal L}}_{GB}=24e^{-2\zt \Phi }(A^{'}+\frac{1}{2}\zt \Phi ^{'})^{2}
[4A^{''}+2\zt \Phi ^{''}+5(A^{'})^{2}+\frac{1}{4}\zt ^{2}
(\Phi ^{'})^{2}+3\zt A^{'}\Phi ^{'}]\nn \\
\bar{{\cal L}}_{(3)}
=-24e^{-3\zt \Phi }(A^{'}+\frac{1}{2}\zt \Phi ^{'})^{4}
[4A^{''}+2\zt \Phi ^{''}+15(A^{'})^{2}+\frac{11}{4}\zt ^{2}
(\Phi ^{'})^{2}+13\zt A^{'}\Phi ^{'}].
\eea
We will eventually resort to $D=5$ dimensions, so
$\zt =4/3$.

The bulk action, in the Einstein frame, now becomes
\bea
S_{bulk}^{(Einstein)}
=\frac{M_{s}^{3}}{2}\int d^{D}x\sqrt{-g}
\left\{\frac{}{}R-\zt (\nabla \Phi )^{2}+\right.\nn \\
+\lm _{0}(\al^{'})e^{-\zt \Phi }
[{\cal L}_{GB}+c_{2}\zt ^{2}(\nabla \Phi )^{4}]+\nn \\
+\lm _{1}(\al^{'})^{2}e^{-2\zt \Phi }
[{\cal L}_{(3)}+c_{3}\zt ^{3}(\nabla \Phi )^{6}]-\nn \\
-\frac{2(D-10)}{3\al ^{'}}e^{\zt \Phi }
\left.+{\cal O}(\al ^{'3})\frac{}{}\right\}.
\eea
Here $c_{2}=(D-4)/(D-2)$ (\cite{mavr1}, \cite{gross}) is the
coefficient of the fourth order dilaton derivative, which is
required by string theory arguments. We assume that the same
is true for the third order term and so $c_{3}$ is a D-dependent
constant. Its precise value would in principle stem from string theory,
however the results presented here do not depend essentially on
its exact value.

Finally, on the brane one can in general consider both the induced gravity
term, a higher order Gauss-Bonnet term and the four-dimensional cosmological
term
\be
S_{brane}=M_{s}^{3}\int d^{4}x\sqrt{-h}\{
^{(4)}R+b_{0}(\al ^{'}) ^{(4)}R_{GB}^{2}-\lm (\Phi )\}.
\ee
Here $b_{0}$ is a dimensionless parameter, as required from
dimensional arguments. For the case of the
conformally flat metric in Eq. (6) the first two terms cancel. In general
however the induced gravity terms for a generic four-dimensional metric
can yield interesting consequences for the junction conditions and for
the behaviour of the bulk solution very close to the brane. This possibility
requires analysis beyond the scope of this work.
The  $\lm (\Phi )=(T/M_{s}^{3})e^{\chi \Phi }$ and
$\chi =5/3$ for a Dirichlet brane, while $\chi =2/3$
for a Neveu-Schwarz brane. Also a gravitational wave
on the brane can for example be considered \cite{char1}, \cite{me}.
\section{The Equations of Motion}\
Substituting Eq. (7) into Eq. (5) and taking care to
include the terms $(\nabla \Phi )^{4},\; (\nabla \Phi )^{6}$
along with Eq. (9), the
total action, in the Einstein frame, becomes
\bea
S_{tot}^{(Einstein)}=\frac{M_{s}^{3}}{2}
\int \int d^{4}xdye^{4A}
\left\{\frac{}{} \right.
&-&8A^{''}-\frac{16}{3}\Phi ^{''}
-\frac{64}{3}A^{'}\Phi ^{'}
-\frac{4}{3}(\Phi ^{'})^{2}-20(A^{'})^{2}+\nn \\
&+&\lm _{0}(\al ^{'})[e^{\zt \Phi }\bar{{\cal L}}_{GB}
+c_{2}\zt ^{2}e^{-\zt \Phi }
(\Phi ^{'})^{4}]+\nn \\
&+&\lm _{1}(\al ^{'})^{2}
[e^{\zt \Phi }\bar{{\cal L}}_{(3)}+c_{3}\zt ^{3}e^{-2\zt \Phi }
(\Phi ^{'})^{6}]+\nn \\
&-&2\Lm e^{\zt \Phi }+{\cal O}(\al ^{'3})+\nn \\
&+&2^{(4)}R\delta (y)-2\lm (\Phi )\delta (y)
\left.\frac{}{}\right\}.
\eea
This is the obvious generalization of the action used in
\cite{binet}.\\
Variation is a tedious but straightforward procedure.
Varying with respect to $A$ gives
\bea
6A^{''}&+&12(A^{'})^{2}+\zt (\Phi ^{'})^{2}+2\Lm e^{\zt \Phi }
+2\lm (\Phi )\delta (y)-\nn \\
&-&\lm _{0}(\al ^{'})e^{-\zt \Phi }
\left[\frac{}{}c_{2}\zt ^{2}(\Phi ^{'})^{4}+\right.\nn \\
&+&24A^{'}(
A^{'}A^{''}+(A^{'})^{3}-2\zt \Phi ^{'}A^{''}
-\zt A^{'}\Phi ^{''}-3\zt \Phi ^{'}(A^{'})^{2}+\zt ^{2}A^{'}
\left.(\Phi ^{'})^{2}\right)\left.\frac{}{}\right]-\nn \\
&-&\lm _{1}(\al ^{'})^{2}e^{-2\zt \Phi }
\left[\frac{}{}c_{3}\zt ^{3}(\Phi ^{'})^{6}+G_{(3)}\right]=0.
\eea
When $\lm _{1}=0$ it reduces to Eq. (14) of \cite{binet}.
Varying with respect to $\Phi $ we exactly obtain
\bea
2\zt \Phi ^{''}&+&8\zt \Phi ^{'}A^{'}-2\Lm \zt e^{\zt \Phi }
-2\frac{d\lm (\Phi )}{d\Phi }\delta (y)-\nn \\
&-&\lm _{0}(\al ^{'})e^{-\zt \Phi }
\left[\frac{}{}c_{2}\zt ^{2}\right.
(\Phi ^{'})^{2}(12\Phi ^{''}+16\Phi ^{'}A^{'}
\left.-3\zt (\Phi ^{'})^{2}\right)+\nn \\
&+&24\zt (A^{'})^{2}
\left(4A^{''}+5(A^{'})^{2}\right)\left.\frac{}{}\right]-\nn \\
&-&\lm _{1}(\al ^{'})^{2}e^{-2\zt \Phi }
\left[\frac{}{}c_{3}\zt ^{3}
(\Phi ^{'})^{4}(30\Phi ^{''}+24\Phi ^{'}A^{'}-10\zt (\Phi ^{'})^{2})
+F_{(3)}\right]=0.
\eea
When $\lm _{1}=0$ it reduces to Eq. (16) of \cite{binet}.\\
The precise form of the
functions $G_{(3)}=G_{(3)}(A^{'},\Phi ^{'})$ and
$F_{(3)}=F_{(3)}(A^{'},\Phi ^{'})$ is given in Appendix I.\\

If we write the generalized Einstein's equations we
also obtain a first order equation, which is nothing more than the
$yy-$component and it is the constraint equation. This
is given by
\bea
12(A^{'})^{2}&-&\zt (\Phi ^{'})^{2}+2\Lm e^{\zt \Phi }+\nn \\
&+&\lm _{0}(\al ^{'})e^{-\zt \Phi }
\left[3c_{2}\zt ^{2}(\Phi ^{'})^{4}-24(A^{'})^{3}\right.
\left.(A^{'}-4\zt \Phi ^{'})\right]+\nn \\
&+&\lm _{1}(\al ^{'})^{2}e^{-2\zt \Phi }
\left[5c_{3}\zt ^{3}(\Phi ^{'})^{6}+H_{(3)}\right]=0,
\eea
where the function $H_{(3)}$ is
\bea
H_{(3)}(A^{'},\Phi ^{'})
=&-&24\cdot 59(A^{'})^{6}-24\cdot 183\zt (A^{'})^{5}\Phi ^{'}
-54\cdot 105\zt ^{2}(A^{'})^{4}(\Phi ^{'})^{2}-\nn \\
&-&30\cdot 130\zt ^{3}(A^{'})^{3}(\Phi ^{'})^{3}
-\frac{15\cdot 201}{2}\zt ^{4}(A^{'})^{2}(\Phi ^{'})^{4}
+\frac{63\cdot 113}{10}\zt ^{5}(A^{'})(\Phi ^{'})^{5}-\nn \\
&-&\frac{71\cdot 303}{40}\zt ^{6}(\Phi ^{'})^{6}.
\eea
When $\lm _{1}=0$ this reduces to Eq. (15) of \cite{binet}.

Taking an appropriate combination of Eqs. (11), (12) and
(13) we obtain, after integration, the first order expression
\bea
e^{4A}\left(\frac{}{}2\zt \Phi ^{'}\right.+3\zt A^{'}&-&4\lm _{0}(\al ^{'})
e^{-\zt \Phi }[c_{2}\zt ^{2}(\Phi ^{'})^{3}-3\zt ^{2}(A^{'})^{2}\Phi ^{'}
+9\zt (A^{'})^{3}]-\nn \\
&-&\lm _{1}(\al ^{'})^{2}e^{-2\zt \Phi }
[6c_{3}\zt ^{3}(\Phi ^{'})^{5}+\left.\frac{}{}I_{(3)}]\right)
={\cal C},
\eea
where
\bea
I_{(3)}(A^{'},\Phi ^{'})
&=&-18\cdot 37\zt (A^{'})^{5}-9\cdot 191
\zt ^{2}(A^{'})^{4}\Phi ^{'}-9\cdot 197
\zt ^{3}(A^{'})^{3}(\Phi ^{'})^{2}-\nn \\
&-&\frac{9\cdot 203}{2}\zt ^{4}(A^{'})^{2}(\Phi ^{'})^{3}
-\frac{9\cdot 209}{8}\zt ^{5}(A^{'})(\Phi ^{'})^{4}
-\frac{63\cdot 193}{20}\zt ^{6}(\Phi ^{'})^{5},
\eea
and ${\cal C}$ is an integration constant.
When $\lm _{1}=0$ it reduces to Eq. (17) of \cite{binet}.\\
Finally we obtain the junction conditions
for a $Z_{2}-$symmetric brane configuration located at $y=0$.
They can be directly deduced from Eqs. (11) and (12),
by integrating on an $\eps-$interval around the brane.
The second derivatives in the first brackets
of the cubic terms (i.e., terms containing $A^{''}$ in $F_{(3)},\; G_{(3)}$)
are writen as total derivatives and after partial integration give
a term on the brane and terms in the bulk that {\it exactly} cancel the
terms in the second brackets (i.e., those containing $\Phi ^{''}$).
Therefore we have
\bea
\left[\frac{}{}3A^{'}\right.&+&4\lm _{0}(\al ^{'})e^{-\zt \Phi }
(3\zt \Phi ^{'}(A^{'})^{2}-(A^{'})^{3})+\nn \\
&+&9\lm _{1}(\al ^{'})^{2}
\left.e^{-2\zt \Phi }J_{(1)}\frac{}{}\right]^{+}_{-}=-\lm (\Phi ),
\eea
where
\bea
J_{(1)}=&-&\frac{2\cdot 59}{5}(A^{'})^{5}-61\zt
(A^{'})^{4}\Phi ^{'}-63\zt ^{2}(A^{'})^{3}(\Phi ^{'})^{2}
-\nn \\
&-&\frac{65}{2}\zt ^{3}(A^{'})^{2}(\Phi ^{'})^{3}
-\frac{67}{8}\zt ^{4}A^{'}(\Phi ^{'})^{4}
-\frac{69}{80}\zt ^{5}(\Phi ^{'})^{5}.
\eea
Also
\bea
\left[\frac{}{}\zt \Phi \right.&-&
2\lm _{0}(\al ^{'})e^{-\zt \Phi }(c_{2}\zt ^{2}(\Phi ^{'})^{3}
+8\zt (A^{'})^{3})-\nn \\
&-&3\lm _{1}(\al ^{'})^{2}e^{-2\zt \Phi }
\left.(c_{3}\zt ^{3}(\Phi ^{'})^{5}+J_{(2)})\frac{}{}\right]^{+}_{-}
=\frac{d\lm (\Phi )}{d\Phi },
\eea
with
\bea
J_{(2)}&=&\frac{6\cdot 122}{5}\zt (A^{'})^{5}
+6\cdot 63\zt ^{2}(A^{'})^{4}(\Phi ^{'})
+6\cdot 65\zt ^{3}(A^{'})^{3}(\Phi ^{'})^{2}+\nn \\
&+&3\cdot 67\zt ^{4}(A^{'})^{2}(\Phi ^{'})^{3}
+\frac{3\cdot 69}{4}\zt ^{5}A^{'}(\Phi ^{'})^{4}
+\frac{3\cdot 71}{40}\zt ^{6}(\Phi ^{'})^{5}.
\eea

\section{Logarithmic Classes of Solutions in the Bulk}\
We consider Eqs. (13) and (15), taking first ${\cal C}=0$
and use the ansatz
\be
A(y)=A_{0}+xln\left(1+\frac{|y|}{y_{*}}\right)
\; \; \;
\Phi (y)=\Phi _{0}-\frac{2}{\zt }
ln\left(1+\frac{|y|}{y_{*}}\right).
\ee
If we had taken ${\cal C}\neq 0$ we would be forced to set
$x=1/4$. The solutions to be presented here are genuine
${\cal O}(\al ^{'2})$-classical solutions, without smooth Einstein
limit.\\
We abbreviate $\al :=\lm _{0}(\al ^{'})$ and
$\tilde{\al }^{2}:=\lm _{1}(\al ^{'})^{2}$.
Then we obtain, from Eq. (15), the constant $y_{*}$ in terms of
the integration constant $\Phi _{0}$ as
\bea
\frac{4\al }{e^{\zt \Phi _{0}}y_{*}^{2}}=
\frac{2(4-3\zt x)\zt ^{3}}
{(8c_{2}\zt ^{2}-9\zt ^{4}x^{3}-6\zt ^{4}x^{2})}
\frac{1}{[1\pm \sqrt{\Dl }]},
\eea
where
\bea
\Dl (x):=1+\frac{\tilde{\al }^{2}}{4\al ^{2}}
\frac{\zt ^{6}(-4+3\zt x)}
{(8c_{2}\zt ^{2}-9\zt ^{4}x^{3}-6\zt ^{4}x^{2})}
\left[-\frac{192c_{3}}{\zt ^{2}}+I_{(3)}(x,-\frac{2}{\zt })\right].
\eea
Since $\Phi _{0}$ must be real, this will restrict the values of x,
as follows below.\\
Also the bulk cosmological constant is given, from Eq. (13), by
\bea
2\al \Lm (x)&=&\left(\frac{4\al }{e^{\zt \Phi _{0}}y_{*}^{2}}\right)
\left(\frac{1}{\zt }-3x^{2}\right)+
\left(\frac{4\al }{e^{\zt \Phi _{0}}y_{*}^{2}}\right)^{2}
\left[-\frac{3c_{2}}{\zt ^{2}}+\frac{3}{2}x^{4}+12x^{3}\right]+
\nn \\
&+&\left(\frac{4\al }{e^{\zt \Phi _{0}}y_{*}^{2}}\right)^{3}
\left(\frac{\tilde{\al }}{\al }\right)^{2}
\left[-\frac{5c_{3}}{\zt ^{3}}\right.
\left.-\frac{1}{64}H_{(3)}(x,-\frac{2}{\zt })\right].
\eea
One can obtain the effective Planck mass up to second order
string corrections in a standard manner. If we had consider
instead of Eq. (6) the metric
\be
ds^{2}=e^{2A(y)}g^{(4)}_{\mu \nu }dx^{\mu }dx^{\nu }+dy^{2}
\ee
and had substituted
\be
R_{\mu \nu \kp \rho }=e^{2A} R^{(4)}_{\mu \nu \kp \rho }
+e^{4A}(A^{'})^{2}(g^{(4)}_{\mu \rho }g^{(4)}_{\nu \kp }
-g^{(4)}_{\nu \rho }g^{(4)}_{\mu \kp }),
\ee
into the Gauss-Bonnet and third order terms, the integrated
coefficient of the four-dimensional scalar curvature $R^{(4)}$
(which is linear in the second derivatives of the metric tensor
function $A(y)$!)
would give the higher order contributions to the effective Planck
mass, as perceived by a four dimensional observer
\cite{jak1}, \cite{mavr2}. Doing this in a careful manner we obtain, up to
${\cal O}(\al ^{'2})$ order,
\bea
M_{Pl}^{2}=M_{s}^{3}\int_{0}^{y_{c}}dye^{2A(y)}
\left[\frac{}{}\right.
1&-&4\al e^{-\zt \Phi }(2A^{''}+3(A^{'})^{2})+\nn \\
&+&48\tilde{\al }^{2}e^{-2\zt \Phi }(A^{'})^{2}
\left.(12A^{''}+11(A^{'})^{2})\frac{}{}\right].
\eea
It follows that
\bea
M_{Pl}^{2}&=&M_{s}^{3}\frac{y_{*}}{(2x+1)}
\left[\left(1+\frac{y}{y_{*}}\right)^{2x+1}\right]_{0}^{y_{c}}
\cdot \nn \\
&\cdot &\left[1-\frac{4\al }{y_{*}^{2}e^{\zt \Phi _{0}}}\right.
(-2x+3x^{2})+
\frac{48\tilde{\al }^{2}}{y_{*}^{4}e^{2\zt \Phi _{0}}}
\left.x^{2}(-12x+11x^{2})\frac{}{}\right].
\eea
We therefore see that when $x\geq -(1/2)$ we can have a finite
effective Planck mass if $y_{*}<0$, so that $y_{c}=|y_{*}|$
and we have a compact proper distance \cite{mavr2}.
On the other hand when $y_{*}<0$ the curvature scalar
$R=-8A^{''}-20(A^{'})^{2}\propto (|y|+y_{*})^{-2}$ diverges as
$y\rightarrow y_{*}$. So we must choose $y_{*}>0$, for an
infinite $y-$direction without curvature singularities
and $x\geq -(1/2)$ in order to assure finite Planck mass.

We now choose $\zt =(4/3), c_{2}=(1/3)$ and compute the term
$I_{(3)}(x,-2/\zt )$. Then
\be
\frac{4\al }{y_{*}^{2}e^{4\Phi _{0}/3}}=
\frac{4(1-x)}{(1-6x^{3}-4x^{2})}
\left[\frac{1}{1\pm \sqrt{\Delta }}\right],
\ee
where
\be
\Delta =1+\frac{\tilde{\al }^{2}}{\al ^{2}}
\frac{4^{5}(1-x)}{3^{2}(1-6x^{3}-4x^{2})}I_{(3)}(x),
\ee
with
\bea
I_{(3)}(x)&=&37x^{5}-191x^{4}+2\cdot 197x^{3}-2\cdot 203x^{2}
\nn \\
&+&209x-\frac{28}{5}\cdot 193+\frac{9}{2}c_{3}.
\eea
If we momentarily switch off the second order corrections
($\tilde{\al }=0$) the allowed values of $x$
(ensuring the positivity of the l.h.s. of Eq. (29))
become $x>1$ or $x\leq 0.4$, as previously noted \cite{binet}.\\
{\bf (+) Branch:} Here we observe the interesting fact that
\be
I_{(3)}^{'}(x)=185(x-1)^{3}\left(x-\frac{209}{185}\right),
\ee
whereas
\bea
I_{(3)}(1)=I_{(3)}(\frac{209}{185})&=&
\frac{9}{2}c_{3}+43-\frac{28}{5}\cdot 193
\nn \\
I_{(3)}(0.4)&=&\frac{9}{2}c_{3}+39.345-\frac{28}{5}\cdot 193.
\eea
So for example, we see that for $c_{3}\geq 231.3$ then
$I_{(3)}(0.4)>0$ and also $I_{(3)}(209/185)>0$. Since
$I_{(3)}^{'}(x)>0$ for $x\geq 209/185$ and for $x\leq 0.4$ we obtain
that $I_{(3)}(x)>0$ and $\Delta (x)>0$. For this range of $x$
therefore, the ${\cal O}(\al ^{'})$ solutions
derived in \cite{binet} are preserved. It is therefore evident that
higher order corrections, in general modify the allowed $x-$range,
due mainly to algebraic constraints.\\
We move now to the general study. The (+) branch gives positive l.h.s.
in Eq. (29) when $\Delta (x)>0$ {\it and}
$x\in (-\infty ,0.4)\cup (1,+\infty )$. The roots of $\Delta (x)=0$
depend also on $\lm _{1}$. Since we deal with the tree level, effective
action for the heterotic string $\lm _{0}=(1/8)$, so
$(\tilde{\al }/\al )^{2}=(\lm _{1}/\lm _{0}^{2})=64\lm _{1}$.
The curves $c_{3}(x)=C(x;\lm _{1})$ (which originate from $\Delta (x)=0$)
on the $(x,c_{3})$-plane,
separate the allowed $x-$regions (where $\Delta (x)>0$) from the forbidden
$x-$regions (where $\Delta (x)<0$), for a given $c_{3}$. We obtain
\bea
\frac{9}{2}c_{3}(x)&=&\frac{3^{2}}{4^{5}}\frac{1}{64\lm _{1}}
\frac{(x-0.4)}{(1-x)}(6x^{2}+6.4x+2.56)
-37x^{5}+\nn \\
&+&191x^{4}-394x^{3}+406x^{2}-209x+\frac{28}{5}\cdot 193.
\eea
In the limit $\lm _{1}\rightarrow 0$ it becomes singular, so what we
present here is a genuine-${\cal O}(\al ^{'2})$ classical solution,
not having a smooth limit to a lower order solution.\\

In Fig. 1 we plot the function $c_{3}(x)$. For the range
of $0.01\leq \lm _{1}\leq 1$ there exists no
significant difference of the plotted curves, as occured from
numerical analysis, so we used for convenience $\lm _{1}=1$.

\begin{figure}[h!]
\centerline{\mbox {\epsfbox{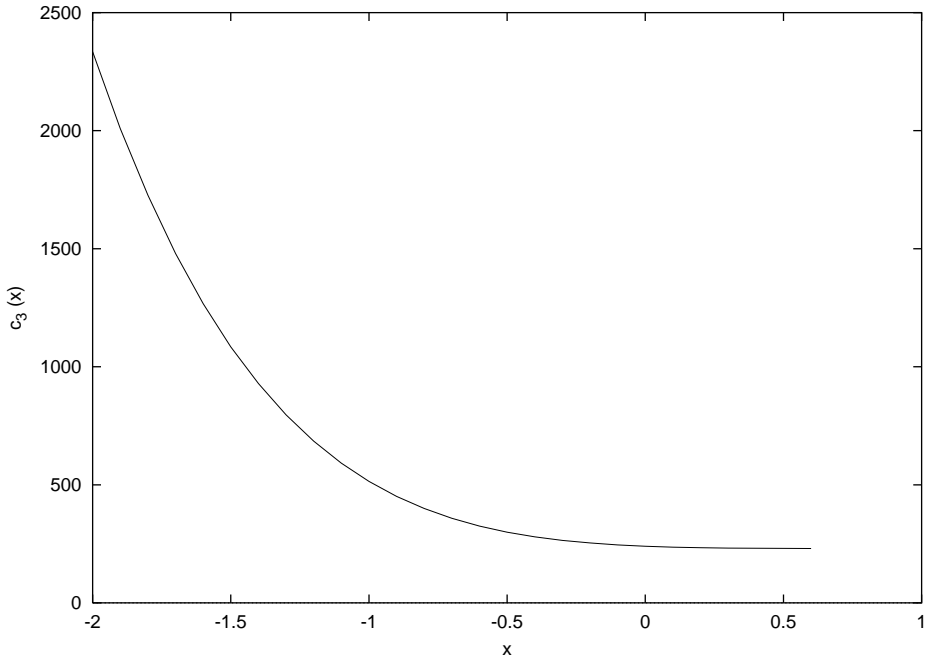}}}
\caption{The dependence
$c_{3}=c_{3}(x)$ for $x\leq 0.4$. The allowed $x-$range is
to the right of the curve, for a given $c_{3}$.}
\end{figure}

As an example for the case of $c_{3}\simeq 500$ the allowed values
of x (those with $\Dl (x)>0$) from precise numerical computation
are $-0.979\leq x\leq 0.4$. Of course we have to add to this the
requirement $-(1/2)\leq x$ which comes from the requirement of finite
effecive Planck mass.\\
In the same
way for $c_{3}\simeq 100$ the allowed values are for $x\geq 2.7719$ from
numerical methods, shown in Fig. 2. As it is evident from the figures,
as $c_{3}$ increases the allowed $x-$range increases.

\begin{figure}[h!]
\centerline{\mbox {\epsfbox{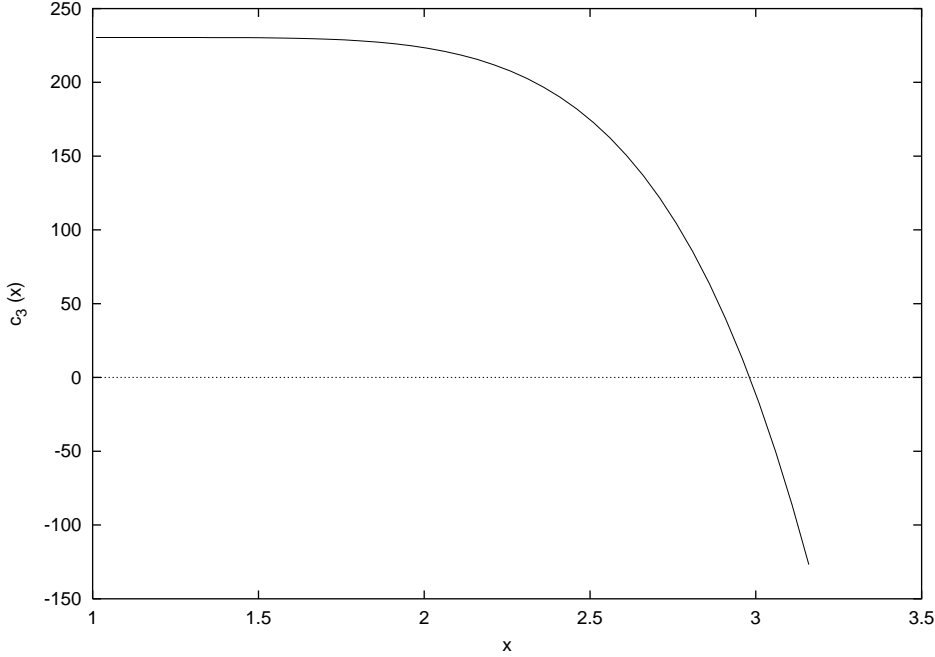}}}
\caption{The function
$c_{3}=c_{3}(x)$ for $x>1.0$.The allowed $x-$region is again
to the right of the curve, for a given $c_{3}$.}
\end{figure}

All the classes of solutions are $\al ^{'}-$solutions, because we
have the additional argument that the cases ${\cal C}=0$ and $\al ^{'}=0$
are incompatible, from Eq. (15).\\

The parameter $x$ is given implicitly, in terms of the bulk
cosmological constant as
\bea
2\al \Lm (x)&=&
\left(\frac{4\al }{y_{*}^{2}e^{4\Phi _{0}/3}}\right)\frac{3}{4}
(1-4x^{2})+
\left(\frac{4\al }{y_{*}^{2}e^{4\Phi _{0}/3}}\right)^{2}
[-\frac{9}{16}+\frac{3}{2}x^{4}+12x^{3}]+\nn \\
&+&64\lm _{1}
\left(\frac{4\al }{y_{*}^{2}e^{4\Phi _{0}/3}}\right)^{3}
[-\frac{135}{64}c_{3}-\frac{1}{64}H_{(3)}(x)],
\eea
with
\bea
H_{(3)}(x)&=&-24\cdot 59x^{6}+48\cdot 183x^{5}-216\cdot 105x^{4}
+\nn \\
&+&240\cdot 130x^{3}-120\cdot 201x^{2}+\frac{63\cdot 113}{5}x
-\nn \\
&-&\frac{568\cdot 303}{5}.
\eea

\begin{figure}[h!]
\centerline{\mbox {\epsfbox{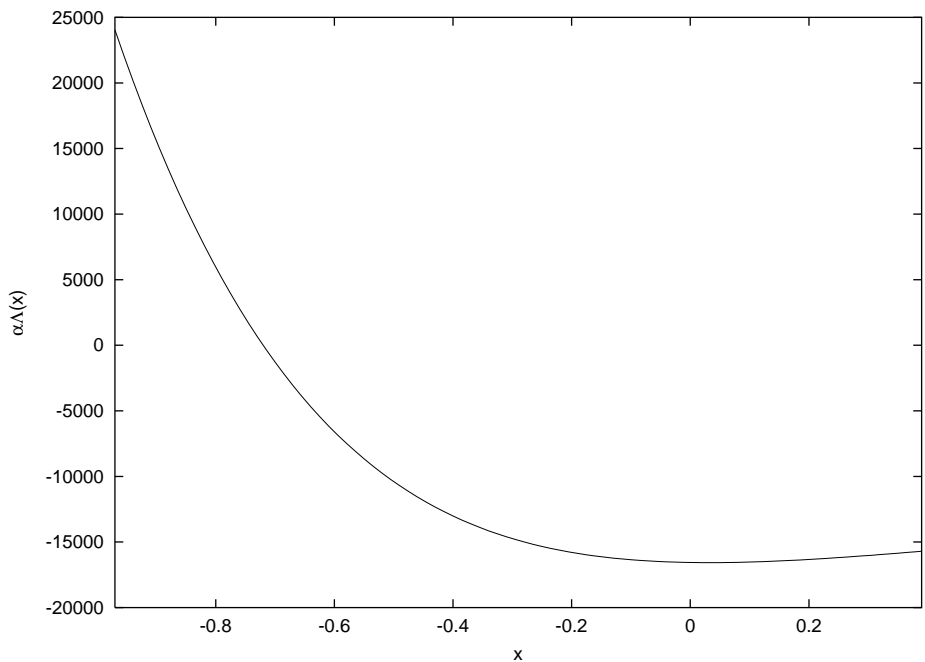}}}
\caption{The bulk Cosmological
constant $\al \Lm (x)$ for $-0.979\leq x\leq 0.4$.}
\end{figure}

In Fig. 3 the bulk cosmological constant for $c_{3}\simeq 500$
and $\lm _{1}=1$ is plotted.

We can also obtain the brane tension, from Eq. (17), as
\bea
\frac{T}{M_{s}^{3}}=-\frac{2}{y_{*}e^{\zt \Phi _{0}/2}}
\left[\frac{}{}\right.3x&+&\frac{4\al }{y_{*}^{2}e^{\zt \Phi _{0}}}
(-6x^{2}-x^{3})+\nn \\
&+&\frac{9\tilde{\al }^{2}}{y_{*}^{4}e^{2\zt \Phi _{0}}}
\left.J_{(1)}(x,-\frac{2}{\zt })\frac{}{}\right],
\eea
while we must have $\chi =\zt /2=2/3$ as required for a Neveu-Schwarz
brane.

Now we are ready to make the order of magnitude estimates that follow drom the
above considerations and constitute our main arguments for this section.
For $x\geq -(1/2)$ up to the allowed value $x\leq 0.4$ and for $c_{3}\simeq 500$,
from Fig. 3 we see that with high precision $2\al \Lm (x)\sim -3\cdot 10^{4}$.
Then using Eq. (36) with $x\simeq 0$ we can estimate the term
\be
l:=(4\al /y_{*}^{2}e^{\zt \Phi _{0}})\simeq 1,
\ee
because we get
$-3\cdot 10^{4}\simeq (3/4)l-(9/16)l^{2}-3.308\cdot 10^{4}l^{3}$.
This is solved for $l\simeq 1$. The same would be obtained for $x\simeq -(1/2)$,
because now we get $-3\cdot 10^{4}\simeq -2l^{2}-2\cdot 10^{4}l^{3}$,
so again $l\simeq 1$. In both cases we have computed the term of
Eq. (37).\\
Now we use $l$ into Eq. (28) with $x\simeq -(1/2)$. We obtain
\be
y_{*}\propto \frac{M_{Pl}^{2}}{M_{s}^{3}}
\frac{1}{(-1+48\cdot 64\cdot 2.1875\lm _{1})}.
\ee
The number 48 comes from the geometric properties of the model, namely
from the contribution of the third order term ${\cal L}_{(3)}$ to the
effective Planck mass. It is therefore expected that higher order
Euler densities when considered in this $\al ^{'}-$expansion, due to the
increased Riemann contractions, will give large numbers. The number 64
comes from the full contribution of these higher order terms
(i.e., when $\lm _{1}=1$) into the string frame effective action, Eq. (5).
Finally the number 2.1875 comes from the increased powers of the $x$ parameter
(namely the brane tension) that will appear as progressively higher order
terms are included in the action \cite{binet}. Also since the allowed $x-$ranges
may include $x-$intervals with large negative numbers, as higher order
terms enter, this number is expected to increase. This is due to
the combination $x^{2}(-12x+11x^{2})$, in Eq. (28). As one includes
progressively higher order terms in the sense described here, and
allows for finite $y-$direction (so the requirement $x\geq -(1/2)$
is relaxed), then allowed $x-$region with large negative values
are expected to contribute with larger factors, as above. \\

The first part of our argument is therefore now
more clear. As we include higher order terms in Eq. (28) the requirement
of $M_{s}\ll M_{Pl}$ appears to be fulfilled more easily, without
severe fine-tuning for $x$, due to the increased contribution of
the term in the second pair of brackets.\\
Now the string coupling, from Eq. (22) is estimated as
$e^{\Phi _{0}}\sim (4\al /y_{*}^{2})^{3/4}$, so
\be g_{s}=
e^{\Phi _{0}}\sim
\frac{M_{s}^{3}}{M_{Pl}^{3}} [-1+6.72\cdot 10^{3}]^{3/2}.
\ee
In \cite{binet} it was correctly observed that when $x\simeq -1/2$
we can have a large string coupling constant $g_{s}=e^{\Phi _{0}}\sim 1$
without having to abandon the requirement of $M_{s}\ll M_{Pl}$.
However this requires a severe fine-tuning of the parameters of
the theory, namely of the brane tension, through $x$. Here however
we see that the inclusion of higher order quantum gravity
correction can induce this result {\it without} severe fine-tuning.
Namely for a whole continuous range of $(-1/2)\leq x\leq 0.4$
we can have $M_{s}\ll M_{Pl}$ while the string
coupling increases as higher order $\al ^{'}$ corrections
are taken into account.\\

So, as it was defined in [29], we consider that the solutions presented
here do not suffer from the strict fine-tuning problem in the following
sense: Namely that
for a whole range of the $x-$parameter which is connected to the brane
tension, through Eq. (38) and the bulk cosmological constant through
Eq. (36), one obtains $M_{s}\ll M_{Pl}$.

In our point of view this is an {\it indirect} support to the claim
that the presence of these higher order quantum gravity corrections in
a sense necessitate the inclusion of higher order {\it quantum
loop-corrections}, {\it as well}, beyond the tree level!
An action originating from non-perturbative quantum gravity theory,
that is not yet availiable, {\it may} solve the fine-tuning problem in
a natural way.\\
Finally from Eq. (38) we have that
\be
T\simeq M_{s}^{4},
\ee
so if we assume that the brane tension
$T\sim (1TeV)^{4}\sim 10 ^{-60}M_{Pl}^{4}$ and the string coupling, from
Eq. (41) still remains small, however several (four) orders of magnitude
larger than the value $g_{s}\sim e^{-45}$ obtained in \cite{binet}.
However as we have just stated
if all the quantum gravity and quantum loop corrections could be included,
in a non-perturbative fashion, even for this, toy-model logarithmic, class
of solutions the need for fine-tuning {\it could} be absent
in order to have $M_{s}\ll M_{Pl}$.

\section{Discussion}\
In this paper we have considered the tree-level, effective action
of the heterotic string in five dimensions with expansion parameter,
in the string frame, the combination $|\al ^{'}R|$. If one tries to
explore, from this point of view, the consequences of incorporating
higher order quantum gravity corrections, then some interesting
facts occur. They can be summarized as follows: Even for the simple, toy
model, class of logarithmic solutions the string scale {\it may} become
much lower than the effective Planck mass without need for severe
fine-tuning of the brane tension. This is due to the increased
contribution of the Riemann
contractions, in the Euler densities. Thus the hierarchy $M_{s}\ll M_{Pl}$
does not depend essentially on the brane tension through the parameter $x$.
Also the string coupling
$g_{s}=exp(\Phi _{0})$ increases by some orders of magnitute for each
contributing term. In \cite{binet} it was pointed that close to the brane,
the string coupling is expected to increase, so it destabilizes the classical
solutions by quantum loop corrections. While this is true, we prefer to
view this as an advantage rather than a disadvantage.
This {\it may point} to the
fact that higher quantum gravity corrections inevitably require also
quantum loop corrections. Thus in the hypothetical non-perturbative
quantum gravity theory the issue of necessity of fine-tuning,
for braneworld solutions, {\it may} be completely absent. In this context the
four-dimensional cosmological constant, through the freedom of its
brane tension component, can be null in an more simple way. The class of
solutions presented here offers, from our point of view an additional
{\it indirect} argument for these claims.

The class of solutions presented here is an extension of solutions that
appeared in [26], [29]. It shares the same basic qualitative behaviour
as those of [29]. From Eq. (28) we have finite effective Planck mass for
$x<-1/2$ namely for the brane tension, if $0<y<y_{c}:=y_{*}$.
Also the spacetime manifold is regular without curvature singularities,
since the curvature scalar $R\propto (|y|+y_{*})^{-2}$ is finite.
However the introduction of the third order term, through $\lm _{1}(\al ^{'})^{2}$,
in the action of Eq. (10) results in a new class of solutions.\\

First this is a genuine ${\cal O}((\al ^{'})^{2})$ class of solutions because
the limit $\lm _{1}\rightarrow 0$ is singular in Eq. (34). It cannot therefore
be mapped to solutions from the lower order approximation. \\

Second the
$x-$interval breaks into allowed regions from Eq. (30) due to the requirement of
$\Dl (x)\geq 0$. For all values of the constant $\lm _{1}\in (0.01,1)$
which controls the contribution of the third order term in the action, Eq. (10),
the curve $\Dl (x)=0$ gives the dependence of the $c_{3}=c_{3}(x)$ on the brane
tension $x$ (Eq. (34) and Figs. 1, 2). This is comparable to the constant
$c_{2}=(D-4)/(D-2)$ which was calculated in [32] and controls the contribution of
the dilaton derivative terms in the action, for the lower order term.
Here we do not have an argument for its exact value. However we found numerically
that for the allowed values of $x<-1/2$, from Fig. 1 and Eqs. (36), (39), that
the relation  $M_{s}\ll M_{Pl}$ through Eq. (28), holds without the need
for an exact fine-tuning for $x$.
So we conjecture that higher order corrections through Euler densities can
satisfy this hierarchy more naturally without the need to fine-tune the brane
tension, as was for example necessary in [29], with $x\simeq -1/2$. \\

Third the string coupling was calculated to increase about four orders
in Eq. (41) through the introduction of the higher order correction. The
string coupling enters at the vertices when loop diagrams are included.
So when it increases one must consider such diagrams in a quantum corrected
action. This {\it may} point to the conclusion that one cannot obtain a
fully consistent picture, unless the succesive quantum gravity corrections
are modified by quantum loop-corrections as well.

Fourth and most important although from Eq. (35) it seems that there
exists a fine-tuning between the bulk cosmological constant and $x$,
inspection of Fig. 3 shows that for the range of $x$ studied here
(namely $-0.5\leq x<0.4$) and the choice of the other parameters,
the bulk cosmological constant assumes an almost {\it constant value},
irrespective of $x$. So the introduction of the third order Euler
density gives a less severe dependence on fine-tuning.

There exists also a second class of solutions with non-zero constant
${\cal C}$, where the relevant set of equations for the metric function $A(y)$
and the dilaton field $\Phi (y)$ admits solutions where the $y-$direction
breaks into two allowed regions, one compact and the other infinite.
In the first case we can have finite effective Planck mass without
curvature singularities, while in the second case one cannot in
general obtain localization of gravity on the brane. These require a detailed
analysis which will appear elsewere.

Finally it would be interesting to consider higher codimensional models
i.e., with action of the form $\int d^{5}xd^{N}X$ \cite{char1} and
in particular of co-dimension two, as for example in \cite{navar}.
In this context the imprints of the cosmological perturbation spectra
are of particular importance in order to decide on the form of the model
that can be viable \cite{cart}. Work along these lines is in progress
\cite{me}.

\newpage

\section{Appendix I}\
The functions appearing in Eqs. (11) and (12) are
\bea
G_{(3)}=36A^{''}\left\{\frac{}{}59(A^{'})^{4}\right.&+&122\zt \Phi ^{'}
(A^{'})^{3}+\frac{189}{2}\zt ^{2}(\Phi ^{'})^{2}(A^{'})^{2}+\nn \\
&+&\frac{65}{2}\zt ^{3}(\Phi ^{'})^{3}A^{'}+
\left.\frac{67}{16}\zt ^{4}(\Phi ^{'})^{4}\right\}+\nn \\
+18\zt \Phi ^{''}\left\{\frac{}{}61(A^{'})^{4}\right.
&+&126\zt \Phi ^{'}(A^{'})^{3}
+\frac{195}{2}\zt ^{2}(\Phi ^{2})^{2}(A^{'})^{2}+\nn \\
&+&\frac{67}{2}\zt ^{3}(\Phi ^{'})^{3}A^{'}
\left.+\frac{69}{16}\zt ^{4}(\Phi ^{'})^{4}\right\}+\nn \\
+24\left\{\frac{}{}59(A^{'})^{6}\right.
&+&111\zt \Phi ^{'}(A^{'})^{5}+
\frac{201}{4}\zt ^{2}(\Phi ^{'})^{2}(A^{'})^{4}-\nn \\
&-&\frac{59}{2}\zt ^{3}(\Phi ^{'})^{3}(A^{'})^{3}
-\frac{579}{16}\zt ^{4}(A^{'})^{2}(\Phi ^{'})^{4}-\nn \\
&-&\frac{201}{16}\zt ^{5}A^{'}(\Phi ^{'})^{5}
\left.-\frac{97}{64}\zt ^{6}(\Phi ^{'})^{6}\right\}
\eea
and
\bea
F_{(3)}=-36\zt A^{''}\left\{\frac{}{}122(A^{'})^{4}\right.
&+&252\zt \Phi ^{'}
(A^{'})^{3}+195\zt ^{2}(\Phi ^{'})^{2}(A^{'})^{2}+\nn \\
&+&67\zt ^{3}(\Phi ^{'})^{3}A^{'}+
\left.\frac{69}{8}\zt ^{4}(\Phi ^{'})^{4}\right\}-\nn \\
-18\zt ^{2}\Phi ^{''}\left\{\frac{}{}126(A^{'})^{4}\right.
&+&260\zt \Phi ^{'}(A^{'})^{3}
+201\zt ^{2}(\Phi ^{2})^{2}(A^{'})^{2}+\nn \\
&+&69\zt ^{3}(\Phi ^{'})^{3}A^{'}
\left.+\frac{71}{8}\zt ^{4}(\Phi ^{'})^{4}\right\}-\nn \\
-24\zt \left\{\frac{}{}170(A^{'})^{6}\right.
&+&378\zt \Phi ^{'}(A^{'})^{5}+
\frac{591}{2}\zt ^{2}(\Phi ^{'})^{2}(A^{'})^{4}+\nn \\
&+&71\zt ^{3}(\Phi ^{'})^{3}(A^{'})^{3}
-\frac{189}{8}\zt ^{4}(A^{'})^{2}(\Phi ^{'})^{4}-\nn \\
&-&\frac{123}{8}\zt ^{5}A^{'}(\Phi ^{'})^{5}
\left.-\frac{71}{32}\zt ^{6}(\Phi ^{'})^{6}\right\}.
\eea
In five dimensions $\zt =(4/3)$ as before.
Equations (11) and (12) could be used also for direct numerical
study of the model.

\section*{Acknowledgments}\
The author would like to thank the Greek State Scholarships
Foundation (I.K.Y) for postdoctoral financial support,
under contract No 422, during this
work.

\end{document}